\documentclass[preprint]{elsarticle}
\usepackage{bm}
\usepackage{amsfonts,amssymb,amsmath}
\usepackage{amsmath,graphicx}
\usepackage{graphicx,pifont} 
\usepackage{epsfig}
\title{Classical many-body systems with retarded interactions: dynamical irreversibility}
\author[ayz]{A.Yu.~Zakharov} 
\ead{Anatoly.Zakharov@novsu.ru}
\author[maz]{M.A.~Zakharov} 
\ead{ma_zakharov@list.ru}

\cortext[cor1]{Corresponding author}
\fntext[fn1]{This work was partially supported by the Russian Ministry of Science and Education in the framework of the base part of state order (Project number 1755).}
\address{Yaroslav-the-Wise Novgorod State University, Veliky Novgorod, 173003, Russia}
 
\begin{document}
\begin{abstract}
The exact equations of motion for microscopic density of classical many-body system with account of inter-particle retarded interactions are derived. It is shown that interactions retardation leads to irreversible behaviour of many-body systems.  
\end{abstract}


\maketitle
\section{introduction}
There is a well known contradiction between reversible equations of classical mechanics and irreversible behaviour of many-body systems.
In particular, it remains an open question about the possibility to describe  transition of many-particle system to equilibrium without the use of the probabilistic hypotheses. Numerous paradoxes, including the Loschmidt and the Zermelo paradoxes~\cite{Kac,Strien}, really prove the incompatibility of deterministic classical mechanics and probabilistic approach to mechanics. In order to use the probabilistic concepts in classical mechanics, a physical source of random behaviour should be indicated. There is no such sources in deterministic classical mechanics.  In particular, there is no random sources for isolated from external world Gibbsian microcanonical ensemble in statistical mechanics.

However, it is very doubtful that non-equilibrium many-body system isolated from the outside world  does not go into an equilibrium state as long as does not interfere  a random external source. Therefore, we must perform an analysis of unstudied factors that can lead to irreversible behavior of many-body systems. There are at least two such factors. 
\begin{enumerate}
	\item Interactions between (electrically neutral) atoms are due to mutual polarization of atoms and depend on their polarizabilities~\cite{Gran,Craig,Buh1,Buh2}. This means that ``instant'' interaction between atoms depend on relatively fast electronic degrees of freedom of atoms. In addition, there are around vacuum fluctuations of the electromagnetic field, manifested in such well-known effects, as the Casimir effect and the Lamb shift. Therefore, there is an additional time-dependent contribution into the interatomic potentials realized through the vacuum fluctuations of the electromagnetic field~\cite{Buh2}. Thus, in addition to the static part of the interatomic potentials there exists fluctuating in time interaction. Note that even insignificant in quantity fluctuation correction to the interatomic potentials can lead to a qualitative change in the behavior of classical systems due to the instability of the dynamical systems trajectories~\cite{Mart}. Probably, this effect can be taken into account by statistical methods only.
	\item There is the retardation effect of interatomic interactions. Accounting for this effect involves the use of systems of functional differential equations, i.e. differential equations involving unknown functions for different argument values. The differences between the argument values of unknown functions are called argument deviations. The mathematical formulation of problems for functional differential equations is radically different from the problems for ordinary differential equations. Unfortunately, currently the theory of functional differential equations is not well developed. More or less effective methods of the problem functional differential equations solutions are realized mainly in the control theory~\cite{Gil}. The hypothesis that the delay interactions can be the cause of irreversibility expressed by various authors (see~\cite{Beau}, for example).
\end{enumerate}

The purpose of this paper is to consider the effect of interatomic interactions retardation on the dynamics of a classical many-body system. 

Unfortunately, the direct account of retarded interactions in conventional approaches of classical mechanics is very troublesome: it is necessary to write a system of ordinary differential equations of motion for all particles of a many-body system in view of the interactions retardation. It is absolutely immense problem. The fact that the usual forms equations of classical dynamics describe the motion of {\em each particle} of the system. For a many-body system such details are redundant. We are interested in the evolution of the particle distribution in the space as a whole.

For this task, we need to describe a many-body system evolution in terms of the local microscopic density. Such a form of the equations of motion in the classical many-body problem for interactions without delay has been proposed in the papers~\cite{Zakh1,Zakh2} 
\begin{equation}\label{density}
	n(\mathbf{r},t) = \sum_{s=1}^N\ \delta\left(\mathbf{r} - \mathbf{R}_s(t) \right) = \int\, \frac{d\mathbf{k}}{\left( 2\pi\right)^3 }\,e^{i\,\mathbf{k}\,\mathbf{r}} \ \tilde{n}(\mathbf{k},t),
\end{equation}
where $\mathbf{R}_s(t)$ is position of $s$-th particle at time $t$, $N$~is the number of particles in the system, $\tilde{n}(\mathbf{k},t)$ are the collective coordinates~\cite{BP}
\begin{equation}\label{tilden(k,t)}
	\tilde{n}(\mathbf{k},t) = \sum_s e^{-i\,\mathbf{k}\,\mathbf{R}_s(t)}.
\end{equation}
In this formulation, we do not assume any probabilistic interpretation of~$n(\mathbf{r}, t)$. Since the dynamical equations of classical mechanics determine the evolution of the system uniquely, the function $n(\mathbf{r},t)$ is a completely deterministic function.

The aim of the present paper is derivation of exact equation for many particle classical system evolution in terms of the local microscopic density with account of the interatomic potentials retardations. This derivation is based on the Newtonian laws only and does not based on any probabilistic or statistical assumptions and hypotheses. It is shown that the retardation interactions leads to irreversible dynamics of many-particle system.

\section{Equations of motion for microscopic density of a many-body system}

To obtain the equations of motion for the function $n(\mathbf{r},t)$ we need  differentiate this function with respect to time and use the Newtonian laws. Since the Newtonian second law contains the second derivatives with respect to time, we should also be restricted to the second derivatives of $n(\mathbf{r},t)$ with respect to time.

\subsection{First order equation}

Let us define the microscopic density~$n(\mathbf{r},t)$ of a classical many-body system by the relation~(\ref{density}). 
The calculation of the sums of type~$ \sum_s \ f \left (\mathbf {R} _s \right) $, where~$ f \left (\mathbf {R} _s \right) $ are any ``the one-particle'' functions, will perform according to the rule
\begin{equation}\label{sum-over-s}
\sum_s \ f\left( \mathbf{R}_s(t)\right) = \sum_s \int\ f\left( \mathbf{r}\right)\, \delta \left(\mathbf{r} - \mathbf{R}_s(t) \right)\, d\mathbf{r} = \int\, f\left( \mathbf{r}\right)\, n\left(\mathbf{r}, t \right)\, d\mathbf{r}. 
\end{equation}  

The first derivative of~$n\left(\mathbf{r}, t \right)$ with respect to time is related to the instantaneous velocities of particles. After differentiating the local density~$n(\mathbf{r},t)$ with respect to time, we have 
\begin{equation}\label{dn/dt}
\dfrac{\partial n(\mathbf{r},t)}{\partial t} = -i \int\, \frac{d\mathbf{k}}{\left( 2\pi\right)^3 }\,e^{i\,\mathbf{k}\,\mathbf{r}} \ \sum_s e^{-i\,\mathbf{k}\,\mathbf{R}_s(t)} \left( \mathbf{k}\cdot \dfrac{d\mathbf{R}_s(t) }{dt} \right). 
\end{equation}
Using~(\ref{sum-over-s}) and Fourier representation of delta-function, we obtain  
\begin{equation}
\dfrac{\partial n(\mathbf{r},t)}{\partial t} = \ \int\ \left( \dfrac{\partial \delta\left(\mathbf{r} - \mathbf{R} \right)}{\partial \mathbf{R}} \cdot \mathbf{v}\left(\mathbf{R}, t\right)\right)  \, n\left(\mathbf{R},t \right)\, d\mathbf{R} = \ -\nabla\left(\mathbf{v}\left(\mathbf{r}, t \right) \, n\left(\mathbf{r}, t \right) \right).
\end{equation}
This is none other than well known equation of continuity, i.e. a local conservation law of the particles number in the system.
\begin{equation}\label{contin}
\dfrac{\partial n(\mathbf{r},t)}{\partial t} + \nabla\left(\mathbf{v}\left(\mathbf{r},t \right) \, n\left(\mathbf{r}, t \right) \right) =0.
\end{equation}
Let us pass to derivation of a second order equation.

\subsection{Second order equation}

Let us write a second derivative of the function~$ n (\mathbf{r}, t) $ with respect to time
\begin{equation}\label{ddot-n}
\dfrac{\partial^2 n(\mathbf{r},t)}{\partial t^2} = \int\, \frac{d\mathbf{k}}{\left( 2\pi\right)^3 }\,e^{i\,\mathbf{k}\,\mathbf{r}} \ \sum_s e^{-i\,\mathbf{k}\,\mathbf{R}_s(t)}\left\lbrace -  \left(\mathbf{k} \cdot \dot{\mathbf{R}}_s (t) \right)^2 -i\left( \mathbf{k}\cdot \ddot{\mathbf{R}}_s(t)\right)  \right\rbrace. 
\end{equation}
The first summand in the integrand of this expression can be transformed to the following integral
\begin{equation}\label{dot-R}
-\sum_s e^{-i\,\mathbf{k}\,\mathbf{R}_s(t)}\, \left(\mathbf{k} \cdot \dot{\mathbf{R}}_s (t) \right)^2\ = \ -\frac{k^2}{3}\, \int\, e^{-i \mathbf{kR}}\, n(\mathbf{R},t)\,v^2(\mathbf{R},t)\, d\mathbf{R}.
\end{equation}

To calculate the second summand in the integrand~(\ref{ddot-n}), we should evaluate~$\ddot{\mathbf{R}}_s$. According to the Newton's second law, we have:
\begin{equation}\label{ddot-Rs}
	\begin{array}{l}
		{\displaystyle \ddot{\mathbf{R}}_s(t)\ =\ -\frac 1{m}\ \nabla_{\mathbf{R}_s} \left[  \sum_{s'} W\left(\mathbf{R}_s(t) - \mathbf{R}_{s'}\left(t - \tau \right)  \right)  + \varphi\left(\mathbf{R}_s,\, t\right) \right] }\\
		{\displaystyle  =\ -\frac 1{m} \nabla_{\mathbf{R}_s} \left[  \int W\left(\mathbf{R}_s - \mathbf{R}'\right) n\left(\mathbf{R}',t-\tau\left(\left|\mathbf{R}_s - \mathbf{R}' \right|  \right)  \right)\, d\mathbf{R}' +  \varphi\left(\mathbf{R}_s,\, t\right)   \right], }\\
	\end{array}
\end{equation}
where~$W\left(\mathbf{R}_s - \mathbf{R}_{s'}\right)$ is an interaction potential between of \textit{resting} particles located in points~$\mathbf{R}_s$ and $\mathbf{R}_{s'}$, $\tau\left(\left|\mathbf{R}_s - \mathbf{R}' \right|  \right)$~is time retardation between points~$\mathbf{R}_s$ and $\mathbf{R}_{s'}$, $\varphi\left(\mathbf{r},\, t \right) $~is an external field potential, $m$~is a particle mass, . 

Substituting~(\ref{ddot-Rs}) into the second summand in the integrand of~(\ref{ddot-n}), we obtain 
\begin{equation}\label{ddot-R}
\begin{array}{l}
{\displaystyle i\sum_s e^{-i\,\mathbf{k}\,\mathbf{R}_s(t)}\, \left( \mathbf{k}\cdot \ddot{\mathbf{R}}_s(t) \right) }\\
{\displaystyle =\ -\frac{i}{m}\, \int e^{-i \mathbf{k} \mathbf{R}}\, n(\mathbf{R},t) \biggl[  \left(\mathbf{k}\cdot \nabla_{\mathbf{R}} \int W\left(\mathbf{R} - \mathbf{R}' \right) n\left(\mathbf{R}',t - \tau\left(\left|\mathbf{R}_s - \mathbf{R}' \right|  \right)\right)\, d\mathbf{R}'  \right) }\\
{\displaystyle 
 +     \left(\mathbf{k}\cdot \nabla_{\mathbf{R}}\,  \varphi\left(\mathbf{R}, t \right) \right)     \biggr] d\mathbf{R}. }
\end{array}
\end{equation}
Using~(\ref{dot-R}), (\ref{ddot-Rs}), and~(\ref{ddot-n}) leads to following results
\begin{equation}\label{ddot-n1}
\begin{array}{l}
{\displaystyle -\int\, \frac{d\mathbf{k}}{\left( 2\pi\right)^3 }\,e^{i\,\mathbf{k}\,\mathbf{r}} \ \sum_s e^{-i\,\mathbf{k}\,\mathbf{R}_s(t)}\left\lbrace \left(\mathbf{k} \cdot \dot{\mathbf{R}}_s (t) \right)^2 \right\rbrace =\, \frac{1}{3}\int \dfrac{\partial^2 \delta \left(\mathbf{r} - \mathbf{R} \right) }{\partial \mathbf{r}^2}\, n(\mathbf{R},t)\,v^2(\mathbf{R},t)\, d\mathbf{R} }\\
{\displaystyle =\, \frac{1}{3}\Delta\left[ n(\mathbf{r},t)\,v^2(\mathbf{r},t)\right]  }
\end{array}
\end{equation}
($\Delta$~ is the Laplace operator)
and
\begin{equation}\label{ddot-n2}
\begin{array}{l}
{\displaystyle \int\, \frac{d\mathbf{k}}{\left( 2\pi\right)^3 }\,e^{i\,\mathbf{k}\,\mathbf{r}} \ \sum_s e^{-i\,\mathbf{k}\,\mathbf{R}_s(t)}\left\lbrace i\left( \mathbf{k}\cdot \ddot{\mathbf{R}}_s(t)\right)  \right\rbrace }\\
{\displaystyle =\,  - \frac{1}{m} \nabla_{\mathbf{r}}  \left[ n(\mathbf{r},t) \left( \nabla_{\mathbf{r}} \left\lbrace  \int W\left(\mathbf{r} - \mathbf{R} \right) n\left(\mathbf{R},t - \tau\left(\left|\mathbf{R}_s - \mathbf{R}' \right|  \right)\right)\, d\mathbf{R} + \varphi\left(\mathbf{r}, t \right) \right\rbrace  \right) \right]. }\\
\end{array}
\end{equation}

Substituting expressions~(\ref{ddot-n1}), (\ref{ddot-n2}) into~(\ref{ddot-n}), we obtain the basic equation 
\begin{equation}\label{equat-bas}
	\begin{array}{r}
		{\displaystyle \dfrac{\partial^2n(\mathbf{r},t) }{\partial t^2}\, = \, \frac{1}{3}\Delta\left[ n(\mathbf{r},t)\,v^2(\mathbf{r},t)\right]   }\\ \\
		{\displaystyle + \frac{1}{m} \nabla_{\mathbf{r}}  \left[ n(\mathbf{r},t) \left( \nabla_{\mathbf{r}} \left\lbrace  \int W\left(\mathbf{r} - \mathbf{R} \right) n\left(\mathbf{R},t - \tau\left( \left|\mathbf{r} - \mathbf{R} \right| \right)  \right)\, d\mathbf{R} + \varphi\left(\mathbf{r},\, t \right) \right\rbrace  \right) \right]}.
	\end{array}
\end{equation}
The further part of the present paper is devoted to the analysis of this equation and its consequences.

\section{Dynamics of slightly inhomogeneous medium}

Let us consider the simplest limiting case of the equation~(\ref{equat-bas}) --- limit a ``nearly uniform'' medium without external field~($ \varphi \left(\mathbf{r}, \, t \right) \equiv 0$):
\begin{equation}\label{homogen}
	\left\lbrace 
	\begin{array}{l}
		{\displaystyle  	n(\mathbf{r}, t)\, = \, n_0 + n_1(\mathbf{r}, t), \quad n_0 = \mathrm{const}, \quad \left| n_1(\mathbf{r}, t) \right| \ll n_0,   }\\
		{\displaystyle  v^2(\mathbf{r}, t)  = \mathrm{const} = v^2.}
	\end{array}
	\right. 
\end{equation}

After linearization of the equations~(\ref{equat-bas}) with respect to~$n_1(\mathbf{r}, t) $, we have
\begin{equation}\label{lin-n1}
	\dfrac{\partial^2n_1(\mathbf{r},t) }{\partial t^2}\, = \, \frac{v^2}{3}\Delta n_1(\mathbf{r},t)  + \frac{n_0}{m} \Delta \int W\left(\mathbf{r} - \mathbf{R} \right) n_1\left(\mathbf{R},t - \tau\left( \left|\mathbf{r} - \mathbf{R} \right| \right)  \right)\, d\mathbf{R}.
\end{equation}
This linear integro-differential equation is different from the usual hyperbolic wave equation by the presence of an integral term of convolution type, so the solution can be found using the Fourier integral
\begin{equation}\label{Four}
	n_1(\mathbf{r},t) = \int\, \frac{d\mathbf{k}\, d\omega}{\left(2\pi \right)^4 }\, \tilde{n}_1(\mathbf{k},\omega)\, \exp \left[i \left(\mathbf{k\cdot r} -\omega t \right)  \right].
\end{equation}
Hence, we obtain an equation for~$\tilde{n}_1(\mathbf{k},\omega)$
\begin{equation}\label{tilde-n1}
	\begin{array}{r}
		{\displaystyle  \int\, \frac{d\mathbf{k}\, d\omega}{\left(2\pi \right)^4 }\,\tilde{n}_1(\mathbf{k},\omega)\, \exp \left[i \left(\mathbf{k\cdot r} -\omega t \right)  \right] }\\
		{\displaystyle  \times \left\lbrace \omega^2 - \frac{v^2}{3}\,k^2 - \dfrac{n_0}{m}\, k^2 \left[ \int d\mathbf{R}\,  W\left(\mathbf{R}\right)\,  e^{i\omega\tau \left(\left| \mathbf{R}\right|  \right) } e^{i\mathbf{k\cdot R}}\right]    \right\rbrace = 0.  }
	\end{array}
\end{equation}
Thus, the dispersion law~$ \omega \left(\mathbf{k} \right) $ of oscillations in  slightly inhomogeneous medium consisting of particles interacting via retarded potentials, satisfies the following equation
\begin{equation}\label{dispers}
	\omega^2 - \frac{v^2}{3}\,k^2 - \dfrac{n_0}{m}\, k^2 \widetilde{W}_{1}\left(\mathbf{k}, \omega \right)  = 0,
\end{equation}
where 
\begin{equation}\label{tildeW-eff}
	\widetilde{W}_{1}\left(\mathbf{k}, \omega \right) = \int d\mathbf{R}\,  W\left(\mathbf{R}\right)\,  e^{i\omega\tau \left(\left| \mathbf{R}\right|  \right) } e^{i\mathbf{k\cdot R}}
\end{equation}
is Fourier-transform  of {\em complex-valued} effective potential
\begin{equation}\label{W-eff}
W_{1}(\mathbf{r}, \omega) = W(\mathbf{r})\, e^{i\omega\tau \left(\left| \mathbf{r}\right|  \right) }.
\end{equation}

Note that in the absence of interactions delay the dispersion law has the following form
\begin{equation}\label{real-disp}
\omega \left(\mathbf{k} \right)\, = \, \pm k\sqrt{\frac{v^2}{3} + \frac{n_0}{m} \widetilde{W}\left( \mathbf{k} \right) },
\end{equation}
where
\begin{equation}\label{tildeW}
\widetilde{W}\left( \mathbf{k} \right) = \left.\widetilde{W}_{1}\left(\mathbf{k}, \omega \right) \right|_{\omega=0} 
\end{equation}
is the \textit{real-valued} Fourier-transform of the interatomic potential

In general case, the roots of the characteristic equation~(\ref{dispers}) with respect to~$\omega$ are complex-valued functions of the wave vector~$ \mathbf{k} $. This fact leads to appearance of both the damped harmonics and the growing harmonics in~(\ref{Four}). Thus, the imaginary parts of the characteristic equation roots lead to irreversible behavior of solutions of the equation~(\ref{lin-n1}).  

The most essential qualitative consequence of the equation~(\ref{dispers}) is the fact that the function~$\omega(\mathbf{k})$ is a complex-valued function. 
This means that retarded interaction between the particles leads to irreversible behavior of a classical many-body system. Irreversible deterministic behavior of a many-body system takes place without any probabilistic assumptions and hypotheses.

\section{The first non-vanishing correction due to the interaction retardation}

In the case of finite range (or rapidly decreasing at large distances) interatomic potentials it is enough to consider the first term due to interaction retardation
\begin{equation}\label{retard}
	n\left(\mathbf{R},t - \tau\left( \left|\mathbf{r} - \mathbf{R} \right| \right)  \right)\, \simeq\,  n\left(\mathbf{R},t \right) \, - \,  \frac{\partial n\left(\mathbf{R},t \right) }{\partial t}   \tau\left( \left|\mathbf{r} - \mathbf{R} \right| \right).   
\end{equation}

Substitute this expression into the equation~(\ref{lin-n1}) and obtain
\begin{equation}\label{n-2}
	\begin{array}{l}
		{\displaystyle \frac{\partial^2 n_1(\mathbf{r},t)}{\partial t^2} - \frac {v^2} 3 \Delta n_1(\mathbf{r},t) - \frac {n_0} m \Delta \int W(\mathbf{r} - \mathbf{R})\, n_1(\mathbf{R},t) \, d\mathbf{R}  }\\
		{\displaystyle + \frac {n_0} m \Delta \int W(\mathbf{r} - \mathbf{R})  \, \tau\left(\left| \mathbf{r} - \mathbf{R}\right|  \right)\, \frac{\partial n_1(\mathbf{R},t)}{\partial t}  \, d\mathbf{R} \, = \, 0. }
	\end{array}
\end{equation}

The first two terms on the left side of this equation correspond to the conventional hyperbolic wave equation. The third term takes into account the contribution of interatomic forces without regard to the retardation. The last term takes into account the retardation and has a first derivative with respect to time. The form of this term is similar to a frictional force (generally speaking, with alternating friction coefficient).

Note that equation~(\ref{n-2}) is somewhat similar to the well-known hyperbolic heat equation (in other words, the heat waves equation)~\cite{ Joseph1,Joseph2}, which provides a finite speed of heat propagation in contrast to the parabolic heat conduction equation.

Let us represent the solution $n_1(\mathbf{r},t)$ of the equation~(\ref{n-2}) in the form of the Fourier integral
\begin{equation}\label{tilde2}
n_1(\mathbf{r},t) = \int\, \frac{d\mathbf{k}}{\left(2\pi \right)^3}\, \tilde{\tilde{n}}_1(\mathbf{k},t) \, e^{i\mathbf{k\cdot r}}.
\end{equation}
As a result, we have the following equation with respect to~$\tilde{ \tilde{n}}_1 (\mathbf{k},t)$:
\begin{equation}\label{tilde-n}
\frac{\partial^{\,2} \tilde{\tilde{n}}_1(\mathbf{k},t)}{\partial t^2} \,
- \, \frac{n_0}{m}\, k^2\, \widetilde{W}_2(\mathbf{k})\, \frac{\partial \tilde{ \tilde{n}}_1 (\mathbf{k},t)}{\partial t}\, + \, k^2 \left[ \frac{v^2}{3}\, +\, \frac{n_0}{m} \widetilde{W}(\mathbf{k})\right] \tilde{ \tilde{n}}_1 (\mathbf{k},t) = 0,
\end{equation}
where $\tilde{\tilde{n}}_1(\mathbf{k},t)$, $\widetilde{W}(\mathbf{k})$, and $\widetilde{W}_2 (\mathbf{k})$~are the Fourier transforms of the functions $n_1(\mathbf{r},t)$, $W(\mathbf{r})$, and $W(\mathbf{r}) \tau (\left| \mathbf{r}\right| )$, respectively. 

The roots of the corresponding characteristic equation in the linear approximation with respect to $\tau$ have the form
\begin{equation}\label{charact}
\begin{array}{l}
{\displaystyle \lambda_{1,2} =  \frac{n_0}{2m}\, k^2\, \widetilde{W}_2(\mathbf{k})\, \pm i k \sqrt{ \left[ \frac{v^2}{3}\, +\, \frac{n_0}{m} \widetilde{W}(\mathbf{k})\right] - \left[ \frac{n_0}{2m}\, k\, \widetilde{W}_2(\mathbf{k}) \right]^2}}\\
{\displaystyle \approx \frac{n_0}{2m}\, k^2\, \widetilde{W}_2(\mathbf{k})\, \pm i k \sqrt{  \frac{v^2}{3}\, +\, \frac{n_0}{m} \widetilde{W}(\mathbf{k}) }.  }
\end{array}
\end{equation}
Depending on the sign of~$ \widetilde{W}_2 (\mathbf{k})$, the corresponding Fourier harmonics are damped oscillations (for $ \widetilde{W}_2 (\mathbf{k}) <0 $), or divergent oscillations (for $ \widetilde{W}_2 (\mathbf{k})> 0 $).
Thus, even an arbitrarily small retardation of the interactions leads to irreversible behavior of many-particle systems. 

\section{Conclusion}

The paper contains the following results.

\begin{itemize}
\item A new form of the equations of motion of many-particle system in classical mechanics is obtained. Within this approach the desired function is the exact microscopic density as a function of position and time. The effect of the  interactions between particles retardation is taken into account. 
\item Derivation and interpretation of obtained equations of motion does not use any statistical and probabilistic assumptions.
\item It is shown that the retardation of inter-particle interactions in the many-body system leads to irreversibility of the system.
\end{itemize}

The basic equation~(\ref{equat-bas}) is not closed, because it contains two unknown functions: the microscopic density~$n(\mathbf{r},t)$ and velocity distribution function~$\mathbf{v}(\mathbf{r},t)$. The second of these functions under certain conditions can be expressed via the velocity potential~$\psi(\mathbf{r},t)$
\begin{equation}\label{v-pot}
\mathbf{v}(\mathbf{r},t)\, = \, -\nabla\psi(\mathbf{r},t).
\end{equation}
In this case, a closed system of equations of motion for many-body system (in  external fields absence) has the following form
\begin{equation}\label{closed}
\left\lbrace 
	\begin{array}{l}
		{\displaystyle \frac{\partial n(\mathbf{r},t)}{\partial t} - \nabla\left[n\left(\mathbf{r}, t \right)\, \nabla\psi(\mathbf{r},t) \right] =0;}\\ \\
	{\displaystyle \dfrac{\partial^2n(\mathbf{r},t) }{\partial t^2}\, - \, \frac{1}{3}\Delta\left[ n(\mathbf{r},t)\,\left( \nabla\psi(\mathbf{r},t) \right)^2 \right]   }\\
	{\displaystyle - \frac{1}{m} \nabla_{\mathbf{r}}  \left[ n(\mathbf{r},t) \left( \nabla_{\mathbf{r}} \left\lbrace  \int W\left(\mathbf{r} - \mathbf{R} \right) n\left(\mathbf{R},t - \tau\left( \left|\mathbf{r} - \mathbf{R} \right| \right)  \right)\, d\mathbf{R}\right\rbrace  \right) \right]=0.}
	\end{array}
	\right. 
\end{equation}
In general, this system of equations is the subject to study. The continuity equation is invariant under the transformation~$t\rightarrow -t$. The basic equation is not invariant with respect to this transformation. Therefore, it can be argued that the presence of interactions delay leads to irreversible behavior of many-body systems.

In this context, there is a lot of interesting open questions.
\begin{enumerate}
	\item Is the Liouville theorem valid for systems with retarded interactions?
	\item Can the equations of continuum mechanics be derived from Can be derived equations of continuum mechanics (for example, the Euler and Navier-Stokes equations) from the the basic equation~(\ref{equat-bas})?
	\item Are the variational principles valid for the equations with retarded interactions?
	\item How to derive the equilibrium properties of many-particle systems? It is clear that as~$t\to\infty$ all the derivatives with respect to time vanish in the basic equation~(\ref{equat-bas}).  Then the constant~$v^2$ should be replaced by~$ \frac{3\kappa T}{m}$~($\kappa$ is the Boltzmann constant, $T$~is absolute temperature), it follows from elementary kinetic theory. This transformation yields a nonlinear integral-differential equation with respect to the microscopic density~\cite{Zakh1,Zakh2}. But still not quite clear how to calculate the full set of the thermodynamic properties of the many-body systems.
\end{enumerate}

\section{Acknowledgements}
We are grateful to Prof. Ya.I.~Granovsky for useful discussions.

This work was supported by the Russian Ministry of Science and Education in the framework of the base part of state order (Project number 1755).

\end{document}